\documentclass[twocolumn,notitlepage,aps,prb,10pt,superscriptaddress,floatfix,letterpaper,reprint
]{revtex4-2}	

\usepackage{hyperref}
\usepackage[usenames,dvipsnames]{color}
\usepackage{amsmath}
\usepackage[english]{babel}
\usepackage{amssymb}
\usepackage{graphicx}
\usepackage{epstopdf}
\usepackage[normalem]{ulem}
\usepackage{subfigure}
\usepackage{todonotes}
\usepackage{braket}
\usepackage{bbold}
\usepackage{placeins}
\usepackage[para]{threeparttable}
\usepackage{lipsum}
\usepackage{multirow}
\usepackage{bm}

\definecolor{linkcolor}{HTML}{399B03}
\definecolor{urlcolor}{HTML}{399B03}
\hypersetup{pdfstartview=FitH, linkcolor=linkcolor,urlcolor=urlcolor,colorlinks=true}
\hypersetup{
    unicode=false,          
    pdftoolbar=true,        
    pdfmenubar=true,        
    pdffitwindow=false,     
    pdfstartview={FitH},    
    pdfauthor={},         
    colorlinks=true,        
    linkcolor=blue,         
    citecolor=red,          
}
\begin{document}

\title{Exploring Coupled Cluster Green's function as a method for treating system and environment in Green's function embedding methods}
\author{Avijit Shee}
\affiliation{Department of Chemistry, University of Michigan, Ann Arbor, Michigan 48109, USA}
\email{ashee@umich.edu}
\author{Chia-Nan Yeh}
\affiliation{
 Department of Physics, University of Michigan, Ann Arbor, Michigan 48109, USA
}
\author{Dominika Zgid}
\affiliation{Department of Chemistry, University of Michigan, Ann Arbor, Michigan 48109, USA}
\affiliation{
 Department of Physics, University of Michigan, Ann Arbor, Michigan 48109, USA
}
\email{zgid@umich.edu}

\begin{abstract}
Within the self-energy embedding theory (SEET) framework, we study coupled cluster Green's function (GFCC) method in two different contexts: as a method to treat either the system or environment present in the embedding construction.
Our study reveals that when GFCC is used to treat the environment we do not see improvement in total energies in comparison to the coupled cluster method itself. To rationalize this puzzling result, we analyze the performance of GFCC as an impurity solver with a series of transition metal oxides. These studies shed light on strength and weaknesses of such a solver and demonstrate that such a solver gives very accurate results when the size of the impurity is small.
We investigate if it is possible to achieve a systematic accuracy of the embedding solution when we increase the size of the impurity problem. We found that in such a case, the performance of the solver worsens, both in terms of finding the ground state solution of the impurity problem as well as the self-energies produced. We concluded that increasing the rank of GFCC solver is necessary to be able to enlarge impurity problems and achieve a reliable accuracy.
We also have shown that natural orbitals from weakly correlated perturbative methods are better suited than symmetrized atomic orbitals (SAO) when the total energy of the system is the target quantity.

\end{abstract}
\maketitle

\section{Introduction}

     Ab initio simulation of strongly correlated molecules, and solids remains still a challenging problem. For molecular problems, many directions were taken primarily based on wave function theories, most notable among them are the complete active space perturbation theory (CASPT2) \cite{CASPT2_JCP90, CASPT2_JCP92}, the n-electron valence perturbation theory (NEVPT2) \cite{nevpt_JCP2001}. However, general applicability of those methods were often problematic because of various theoretical and numerical challenges. The most severe challenge is the growth of the dimension of the Hilbert space once the number of active orbitals increases, effectively making CASPT2 and NEVPT2 calculations very difficult or impossible when the size of the active space exceeds 18 orbitals and when the full configuration interaction (FCI) is used to treat the active space. Additional challenges appear due to the insufficient treatment of dynamical correlation by the second order perturbation theory, intruder state problems \cite{} present in CASPT2,  or the requirement of higher body density matrices present in both CASPT2 and NEVPT2.
     These above difficulties are the primary reason why CASPT2 or NEVPT2 cannot be used for treatment of molecules containing multiple transition metal atoms or strongly correlated solids.
    
      For material simulation, among the first principle methods the Green's function based dynamical mean field theory (DMFT)~\cite{Georges96,Kotliar06} has achieved a considerable success. DMFT relies on the physical idea that it is possible to capture important physics of a problem by embedding the strongly correlated part of the problem (system) within the remaining orbitals (environment) via a low-energy Hamiltonian. Though conceptually the same idea can be applied for molecular simulation as well, the attempt to apply DMFT-like methods to molecular simulation is quite recent \cite{Millis_PRB11,Zgid_2011,Tran_jcp_2015,Millis_PRB2015,Tran_jctc_2016,Tran_generalized_seet,Tran_Shee_2017}. Apart from studying challenging strong correlation problems, molecular simulation often provides ideal test bed for a particular realization of quantum embedding method \cite{PhysRevX.7.031059,haule_lee_h2_2017}, which is often helpful prior to more demanding material simulations. In this work, we will discuss application of embedding theories to molecules. 
      
      In practical applications of the DMFT method, often the density functional theory (DFT) is chosen as a low level method for the entire system, resulting in the DFT+DMFT~\cite{LDAplusDMFT_Anisimov1997,Kotliar06} method. Though, this approach was applied very successfully for various problems \cite{Kent_science18}, it suffers from two difficulties: a) often the double counting of electron correlation between two different methods used to treat the environment and system is hard to eliminate; b) accurate effective interactions are necessary to describe the the system. In order to circumvent those problems, the self-energy embedding theory (SEET)~\cite{Kananenka15,Zgid17,Rusakov19,Iskakov20}  was proposed, where double counting is eliminated by Feynman diagrammatic construction of the theory, and ab initio interactions were considered among the strongly correlated system's orbitals, thereby avoiding any necessity of effective (often empirical) parameters. SEET was applied successfully to both molecular and periodic systems. In this article we will restrict ourselves only to the molecular problems.
      
      For molecules, strong correlation typically features near bond dissociation, and in excited states of most of the molecules, and in the ground states of transition metal and lanthanide containing molecules. For SEET, similar to the active space methods, one crucial challenge is to identify a set of strongly correlated orbitals. For most of the problems, such identification often relies on ``chemical intuition'', though automatic constructions are also known \cite{pulay_uno_jcp88,khedkar_JCTC20,reiher_autocas_JCC19}. 
      A systematic procedure of increasing the number of active orbitals could essentially alleviate the challenges of the active space choice. In such a procedure, the size of the active space could be systematically increased based on correlated natural orbital (NO) occupations or even based on chemical intuition resulting in  reaching a convergence with respect to the number of active orbitals. 
      For Green's function embedding methods, such as SEET, such an approach would require a robust method that can provide a solution for a large number of strongly correlated orbitals.
      Particularly, one should be able to increase the size of the active space  (or number of treated strongly correlated orbitals) without sacrificing the accuracy of the solution. In SEET, we typically employ exact diagonalization (ED) as a method of choice for solving the strongly correlated part of the problem. However, because of exponential complexity of ED, we cannot systematically increase the active space size. Previously, for this problem truncated CI solvers were proposed \cite{Zgid2012, Zgid_2011}, however, they do not fully remedy the problem of the fast increase of the Hilbert space dimension of the active space.
      As an alternative to ED or truncated CI type solvers for strongly correlated part of the problem, one can consider coupled cluster (CC) based methods. 
      They are likely to be more successful because the convergence of CC with respect to the rank of excitation is much faster \cite{Musial_Bartlett_RMP07} than that for CI based methods, also it does not suffer from the size extensivity error of truncated CI). Consequently, much larger orbital spaces should be accessible to CC based solvers.             
      
      In order to employ CC as a solver for SEET, a Green's function mapping of the CC wave function (GFCC) is required. In our previous work~\cite{Shee2019}, we have implemented such a method using the Lanczos tridiagonalization procedure, which is computationally more efficient than some other related implementations~\cite{KowalskiJCP2014, Zhu_PRB19} because the scaling of that procedure is completely independent of the size of the frequency grid and  avoids solving linear equations at every frequency points, which ultimately may be unstable. Additionally, the same solution of the tridiagonalization procedure may be used to prepare a desired Green's function either on a real or imaginary frequency grid thus avoiding solving a separate set of linear equations for each of the grids.
      GFCC has been employed for Green's function based embedding theories by us and by Zhu \textit{et. al.} \cite{Zhu20_HF_CCSD,Zhu_DMFT_PRX21}. 
      While it is clear that GFCC has many advantages as a solver for either DMFT or SEET, it is significant to note that when it is based on CCSD only a limited set of excitations are included. Consequently, it is important to analyze
       i) how GFCC performs in comparison to an exact solver such as ED for realistic problems; ii) whether GFCC preserves the same level of accuracy with the increasing active space size. 
       The investigation of the last point is especially crucial in understanding if SEET or DMFT employing GFCC can be converged with respect to the active space size.
       In this paper, we will numerically investigate all those aspects with a series of transition metal oxides taken from the Simons benchmark \cite{MB_comparison:2020}, for which we have very good theoretical reference results.
      
      Moreover, SEET allows us to consider GFCC as a low level method used to treat weakly correlated orbitals (environment). In such SEET(CCSD/ED) calculations,  ED is considered as a high-level solver while CCSD is used to illustrate weak correlations. This could be utilized for the multi-reference molecular problems where dynamical correlation plays crucial role. One such example is the avoided curve crossing of lithium fluoride (LiF) molecule. If successful, such an approach could capture similar physical effects as in a multi-reference coupled cluster theory. We will discuss the SEET(CCSD/ED) approach and  analyze reasons why such an approach did not perform well. We base our discussion based on an example molecule from the Simons benchmark~\cite{MB_comparison:2020}.
      
      This paper is organized as follows.
      In Sec. \ref{sec:seet_theory}, we describe the key ingredients of SEET theory relevant for this work. In Sec. \ref{sec:ccsd_nonloc}, the theoretical description and numerical studies are provided when GFCC was used as a non-local method necessary to describe weak correlation. In Sec. \ref{sec:ccsd_local}, we will recapitulate some of the theoretical aspects of GFCC as a solver. The key computational aspects of this work will be described in Sec. \ref{sec:organization}, while different choices of impurity basis will be compared in order to define the impurity problem in Sec. \ref{Sec:imp_choice}. Finally, main results including the transition metal oxides will be shown and analyzed in Sec. \ref{sec:results}. In Sec. \ref{sec:concl}, we will summarize the performance of GFCC both as a solver and as a method for describing the weak correlation in the environment.

\section{Theory} \label{sec:seet_theory}
SEET is a Green's function quantum embedding theory capable of describing systems that require different level of methods to describe various types of electronic correlation present in the system of interest. 
The diagrammatic series that arises in SEET can be viewed as an approximation to the self-energy generated by the Luttinger-Ward functional, $\Phi$, consisting of a set of 
closed, bold, two-particle irreducible Feynman diagrams (also known as ``skeleton'' diagrams) expressed using a fully interacting Green's function $G$ and bare interaction $v$. In SEET, first a low level (usually perturbative) method is used to evaluate a set of self-energy diagrams for all the orbitals present in the system. We will call this part of the diagrammatic self-energy expansion  $\Sigma_{weak}$. Subsequently, a set of  orbitals (creating an ``active space'' part of the system) is identified and a method (usually non-perturbative) capable of treating strong correlation is employed to evaluate a set of self-energy diagrams that provide some correction over the ones evaluated using a weakly correlated method. This part of the diagrammatic expansion is denoted as $\Sigma_{strong}^{i}$, where the superscript $i$ denotes the possibility of multiple orbital groups $i$ that can be treated separately. We denote the total number of them by $N_{imp}$. The addition of both the ``weak'' and ``strong'' parts of the self-energy requires a cancellation of any double counting contributions (here called $\Sigma_{weak}^{i}$) between the methods capable of describing weak and strong correlations. Consequently, we define the self-energy obtained in SEET as $\Sigma^{SEET}$  
\begin{equation}
    \Sigma^{SEET} = \Sigma_{weak}^{total} + \sum_{i=1}^{N_{imp}} \Sigma_{strong}^{i} - \Sigma_{weak}^{i}. 
\end{equation}\label{eq:sigma_assembly}
Note, however, that the above prescription for the SEET self-energy does not specify how its different parts can be evaluated in practical calculations. Since the method for describing weak correlation is usually computationally affordable,  the evaluation of the $\Sigma_{weak}^{total}$ and $\Sigma_{weak}^{i}$ can involve all the orbitals present in the system. However, the method for treating strongly correlated subset of orbitals is computationally expensive and it is customarily applied only to few orbitals that constitute an open system, here expressed as an impurity orbitals coupled to a non-interacting bath. 
For SEET,  we consider only the Fock matrix and bare two-body integrals of the parent system  for the impurity construction. Therefore it is possible to employ multitude of ``quantum chemistry'' methods as impurity solvers formally necessary for evaluation of the $\Sigma_{strong}^{i}$ term.
The above defined construction of the $\Sigma^{SEET}$ self-energy provides a useful mean to approximate the full diagrammatic series of the Luttinger-Ward $\Phi^{LW}$ and the self-energy obtained in it $\Sigma^{LW}$. 

In a similar spirit  GW+(E)DMFT\cite{GWplusEDMFT_Sun02,Biermann03,Boehnke16,Zhu_DMFT_PRX21,multitier_GW+DMFT_werner_2017}, are also commonly used to provide an approximation to the exact functional $\Psi$ of Almbladh \cite{psi_almbladh99} \textit{et al.} (which is a Legendre transform of the $\Phi^{LW}$ functional). However, as opposed to SEET,  $\Psi$ functional is a functional of G and renormalized interaction $W(\omega)$, that is $\Psi(G,W)$. Since $W(\omega)$ is frequency-dependent, at present not many solvers can solve successfully such a defined  impurity problem. 


In SEET and in GW+(E)DMFT, it is customary to write the self-energy taking into consideration a regrouping of terms from Eq.~\ref{eq:sigma_assembly}, resulting in the SEET self-energy expressed as
\begin{equation}
\begin{split}
\Sigma^{SEET} = \Sigma_{weak}^{non-local}
-[\Sigma^{local}_{strong}]^{i}, \\
\Sigma_{weak}^{non-local} = \Sigma_{weak}^{total}  - [\Sigma^{local}_{weak} ]^{i},\\
[\Sigma^{local}_{strong} ]^{i} =  \Sigma_{strong}^{i},
\end{split}
\end{equation}
where the term $\Sigma$ contains both the frequency dependent and independent (static) parts, $\Sigma=\Sigma(\omega)+\Sigma_\infty$.
Operationally, it means that the self-energy can be evaluated at different level of theory, and then assembled  according to Eq.~\ref{eq:sigma_assembly} to get $\Sigma^{SEET}$ for the total system. This  allows us to evaluate $\Sigma^{non-local}$ and $[\Sigma^{local}]^{i}$ at various level of theories. In this work, we explore the GFCCSD method to evaluate  both the $\Sigma^{non-loc}$ and $\Sigma^{loc}$ components of the self-energy. Both cases are elaborated in two subsequent sections.

Here, for the ease of a reference, we list general steps of the SEET algorithm. In the next sections, we will refer to the modification of these steps that are necessary to employ GFCCSD either as a local and non-local method.
Here, we describe the SEET algorithm from a general perspective for molecular problems. For a SEET version for periodic system, please refer to Ref. ~\citenum{Iskakov20}.
In the above description, we denote by \textbf{LL} a low level part of the SEET loop, where we employ a method capable of describing weak correlation.   \textbf{HL} is used to denote a high level method that is employed in a smaller orbital subset for the description of strong correlation.

\begin{description}

\item [LL0] Perform Green's function second order (GF2)~\cite{Dahlen05,Phillips14,Phillips15,Rusakov14,Rusakov16,Welden16,Kananenka_grid_16,kananenka_hybrif_gf2,Iskakov_Chebychev_2018} 
or GW~\cite{Hedin65,Aryasetiawan98,QPGW_Schilfgaarde,Stan06,Koval14,GW100,Holm98,Tran_Shee_2017} for the whole system.

\item [LL1] Evaluate  Green's function, $G(\omega)$, self-energy $\Sigma(\omega)$, Fock, $F$, and  overlap,  $S$ matrices for the system of interest. These quantities involve all orbitals of the system.
\item [LL2] Find the chemical potential $\mu$ to ensure that the number of electrons in the whole system is proper.
\item  [LL3] Find an orthogonal orbital basis and appropriate transformation matrices $C$ that are suitable for the problem of interest.
\item [LL4] Using  $C$ transform all the quantities such as $G(\omega)$,  $\Sigma(\omega)$, and $F$ to the chosen orthogonal basis.
\item [LL5] Prepare one-body Hamiltonian $F^{NDC}$ for the impurity construction. This is a Fock matrix with the double counting correction that is subtracted.
\begin{align}
[F^{NDC}]_{ij}= {}&F_{ij}-[\Sigma_{\infty} ]_{ij} \nonumber \\
[\Sigma_{\infty} ]_{ij, \sigma}= {} &\sum_{kl}{(}\sum_\sigma \gamma_{kl, \sigma}v_{ijkl}{)}- {(}\gamma_{kl, \sigma} v_{ilkj}{)} \label{eq:sigma_infty},
\end{align}
where $\gamma=-G(\tau=\beta)$, is the correlated one-body density matrix evaluated from the correlated Green's function, and $\sigma$ corresponds to the individual spin indices. We will omit the spin index from all other quantities except $\Sigma_\infty$, because they remain the same for both up and down spins. The inverse temperature is denoted as $\beta$. Both labels $i,j$ are restricted to the chosen subset of orbitals (impurity orbitals).

\item [HL1] \label{embedding_loop1} 
In the orbital basis of interest perform an embedding construction. The subset of the Green's function of the total system for which the strong correlation is important is evaluated as 
\begin{eqnarray}
[G(\omega)]_{sub}=\bigg[\Big[(\omega+\mu)\mathbf 1-F-\Sigma_{tot}(\omega)\Big]^{-1}\bigg]_{sub}
\end{eqnarray}
using the quantities ($F$ and $\Sigma_{tot}$) evaluated in the full orbital space.
The total self-energy is defined as $\Sigma_{tot}(\omega)$.
The hybridization, $\Delta(\omega)$, between the chosen orbital subset with the rest of system can be evaluated from the following expression
\begin{eqnarray}\label{hybrid}
[G(\omega)]_{sub}=\Big[(\omega+\mu)\mathbf 1_{sub}-F_{sub}-[\Sigma_{tot}(\omega)]_{sub}-\Delta(\omega)\Big]^{-1}.
\end{eqnarray}
Note that the size of the subset of the strongly correlated orbitals is $n_{act}$ which is the number of active/chosen or impurity orbitals. 
$[\Sigma_{tot}(\omega)]_{sub}$ is the subset of the total self-energy constructed according to Eq.~\ref{tot_se}. 

\item [HL2] For the subset of chosen orbitals prepare an impurity model. 
This requires transforming two-body integrals to the chosen orthogonal basis, however, this transformation has to be performed only for the impurity orbitals.

\item [HL3]  Find the hybridization function from Eq.~\ref{hybrid}. Evaluate $\epsilon_k$ and $V_{ij}$ that yield the  best least-square fit to the evaluated hybridization~\footnote{Note that this step is only necessary when working in an explicit Hamiltonian formulation with a finite bath.} according to the equation
\begin{equation}\label{hybrid_explicit}
\Delta_{ij} (\omega)\approx\sum_{k}^{M}\frac{V_{ik}V_{jk}}{\omega-\epsilon_k}.
\end{equation}
The total number of bath orbitals is denoted as $M$. 
For more information about bath fitting, see Ref.~\citenum{Liebsch_2011}.

\item [HL4] Employ a solver capable of dealing with the non-diagonal hybridizations. In this paper, we employ the exact diagonalization (ED) solver or GFCC 
to find $[\Sigma_\text{strong}(\omega)]_{imp}$. The dimension of this  matrix is $n_{act}\times n_{act} \times n_{\omega}$, where $n_{\omega}$ is the number of frequency grid points.

\item [HL5] Only for the impurity orbitals evaluate the double counting correction $[\Sigma_\text{weak}(\omega)]_{imp}$  using the chosen orthogonal basis.

\item [HL6]  Remove the double counting correction and prepare a new self-energy in the following way 
\begin{eqnarray}\label{se+imp}
[\Sigma(\omega)]_{imp}&=[\Sigma_\text{strong}(\omega)]_{imp}-[\Sigma_\text{weak}(\omega)]_{imp}.
\end{eqnarray}
Note that this total impurity self-energy $[\Sigma(\omega)]_{imp}$ is evaluated in the chosen orthogonal basis.\\

\item [HL7]  Construct the total self-energy of a subset of chosen orbitals as 
\begin{align}\label{tot_se}
[\Sigma(\omega)]_{sub} = {}& [\Sigma(\omega)]_{imp}+[\Sigma_\text{weak}(\omega)]_{sub}\\
={}&[\Sigma_\text{strong}(\omega)]_{imp}+[\Sigma_\text{weak}(\omega)]_{sub}-[\Sigma_\text{weak}(\omega)]_{imp} \\
={}&[\Sigma^{local}_\text{strong}(\omega)]_{imp}+[\Sigma^{non-local}_\text{weak}(\omega)]_{sub}
\end{align}
This subset self-energy contains both the impurity self-energy $[\Sigma_\text{strong}(\omega)]_{imp}$ which we will call an embedded contribution and the ``embedding'' self-energy contribution obtained by a lower level method, here $[\Sigma^{non-local}_\text{weak}(\omega)]_{sub}=[\Sigma_\text{weak}(\omega)]_{sub}-[\Sigma_\text{weak}(\omega)]_{imp}$. 

\item [HL8]  Build a new Green's function using the previously evaluated self-energy
\begin{equation}
G(\omega)=[(\omega+\mu)S-F-\Sigma(\omega)]^{-1}.
\end{equation}
The self-energy matrix is defined in the following way:  for indices $i,j$ belonging to the chosen/active orbitals the total self-energy $\Sigma(\omega)$ is constructed according Eq.~\ref{tot_se}. For the remaining orbitals ($i,j\notin sub$), the self-energy  should be constructed as
\begin{equation}
[\Sigma(\omega)]_{pq \notin sub}=[\Sigma_\text{weak}(\omega)]_{pq \notin sub},
\end{equation}
\begin{equation}
[\Sigma(\omega)]_{m \in sub \ n \notin sub}=[\Sigma_\text{weak}(\omega)]_{m \in sub \ n \notin sub}.
\end{equation}

\item [HL9]  Find a new chemical potential $\mu$ to ensure a proper number of electrons in the whole system. Evaluate the one-body density matrix.
\item [HL10] 
Continue with the update of the hybridization and iterations starting from the step~{\bf HL3}. Check for the convergence of the impurity self-energy in each of the iterations.

\item [LL5] Go to point~{\bf LL0} performing only a single iteration of GF2 or GW using a correlated one-body density matrix evaluated  in point~{\bf HL9}.
\end{description}

\section{CCSD as a non-local method for SEET} \label{sec:ccsd_nonloc}
CCSD is a zero-temperature formalism based on Goldstone diagrams, where we do not have an explicit diagrammatic expansion of the  CCSD self-energy.   
However, it is possible to map CCSD wave function to a Green's function (G$^{CCSD}$) via the Lehmann representation, and express the self-energy ($\Sigma^{CCSD}$) in terms of the Dyson equation
\begin{align}\label{eq:ccsd_sigma}
    \Sigma^{CCSD} = {}& [G^{HF}]^{-1} - [G^{CCSD}]^{-1} \\
                  = {}& (\omega+ \mu) - F^{HF} - [G^{CCSD}]^{-1},
\end{align}
where $G^{HF}$ and $F^{HF}$ are Hartree-Fock Green's function and Fock matrix, respectively. 

Since $\Sigma^{CCSD}(\omega)$ is explicitly evaluated using a Hatree-Fock reference, the static part of the self-energy 
contains a spurious constant contribution, $\Sigma^c$ that arises due to a lack of self-consistency affecting the static part of the self-energy
\begin{equation}
   \Sigma^{CCSD}(\omega) = \Sigma_{\infty} + \Sigma(\omega) + \Sigma^c,
\end{equation}{}
where, $\Sigma_{\infty}$ is evaluated using Eq. \ref{eq:sigma_infty} with full set of 2-electron integrals, $v$, and $\gamma = -G^{HF}(\tau=\beta)$.
As a result, the use of $\Sigma^{CCSD}$ could be compared to the self-energy arising in one-step GF2 and GW methods.

In the following section, we will delineate the steps involved in a SEET(CCSD/ED) calculation, where CCSD is used as a low level method for treating weak correlations, while ED is employed to deal with strongly correlated orbitals.

\begin{description}
\item[LL0]  At zero temperature, for all orbitals present in the system of interest, perform CCSD calculation.

\item[LL1] Calculate the CCSD Green's function starting from amplitudes T$_1$ (singles), T$_2$ (doubles) for the right hand CC wave function, and $\Lambda_1$ and $\Lambda_2$ for the left hand wave function. 
The CCSD Green's function is calculated in the molecular orbital (MO) basis.
\item[LL2] Adjust the chemical potential $\mu$ to get a correct number of electrons in the system. 
\item[LL3] Transform Green's function from the MO basis to the atomic orbital (AO) basis and calculate density matrix and natural orbitals. 
\item[LL4] Identify the natural orbitals with significant occupations and construct impurities for these orbitals. These natural orbitals serve as a basis for impurity solver.
The Green's functions, Fock matrix $F$, the one-body density matrix, $\gamma$ and two-body integrals, $v$ are transformed to the NO basis, allowing us to calculate the CCSD self-energy directly in the NO basis according to the Dyson equation (Eq.~\ref{eq:ccsd_sigma}).
Here $\Sigma^{CCSD}=\Sigma^{total}_{weak}$


\item[LL5] Subtract any local contribution (double counting) due to CCSD being used as a weak correlation method. $\Sigma_{\infty}^{DC}$ is calculated in the following way:
\begin{equation}
[\Sigma_{\infty} ]_{ij, \sigma}= \sum_{kl}{(}\sum_\sigma \gamma_{kl, \sigma}^{HF} v_{ijkl}{)}- {(}\gamma_{kl, \sigma}^{HF} v_{ilkj}{)},
\end{equation} \label{Eq:staticdc}

where, $i,j,k,l,\dots$ denote the impurity orbitals. However, note that here the one-body density matrix comes from an earlier Hartree-Fock calculation rather than a correlated calculation. This is only the static double counting part of the self-energy. The calculation of the dynamic double counting part of the CCSD self-energy $[\Sigma_{weak}^{CCSD}]_{imp}$ requires a different treatment than $[\Sigma_{weak}^{GF2}]_{imp}$ or $[\Sigma_{weak}^{GW}]_{imp}$ since these can be directly evaluated from the diagrammatic expressions for the self-energy.
Consequently, $[\Sigma_{weak}^{CCSD}]_{imp}$ has to be evaluated explicitly by solving the impurity problem.

\item[HL1-HL3] Same as in the general algorithm.
\item[HL4] Use an ED solver to evaluate  $[\Sigma_{strong}^{ED}]_{imp}$ and CCSD to evaluate $[\Sigma_{weak}^{CCSD}]_{imp}$.
Both these self-energies are evaluated using the Dyson equation in the following way
\begin{align}
    [\Sigma]_{imp} = {}& [G_{0}]^{-1} - G_{imp}^{-1} \\               = {}& \omega - F^{NDC} - \Delta(\omega) - G_{imp}^{-1},   
\end{align}
where $G_{imp}$ is the impurity Green's function calculated either using ED when $[\Sigma_{strong}^{ED}]_{imp}$ or with CCSD when $[\Sigma_{weak}^{CCSD}]_{imp}$ is evaluated.

\item[HL5] Save $[\Sigma_{weak}^{CCSD}]_{imp}$ as the frequency dependent double counting correction.

\item[HL6-HL7] Construct the total self-energy as 
\begin{equation}
    [\Sigma(\omega)]_{sub} = 
    [\Sigma_{strong}^{ED}]_{imp}-[\Sigma_{weak}^{CCSD}]_{imp}+[\Sigma_{weak}^{CCSD}]_{sub}.
    \label{Eq:sigma_update_cc-nonloc}
\end{equation}

\item[HL8-HL10] Same as in the previous algorithm.
\item[LL6] Same as in the previous algorithm.


    


\end{description}{}

\begin{figure}
\begin{center}
\includegraphics[width=\columnwidth, height=8cm]{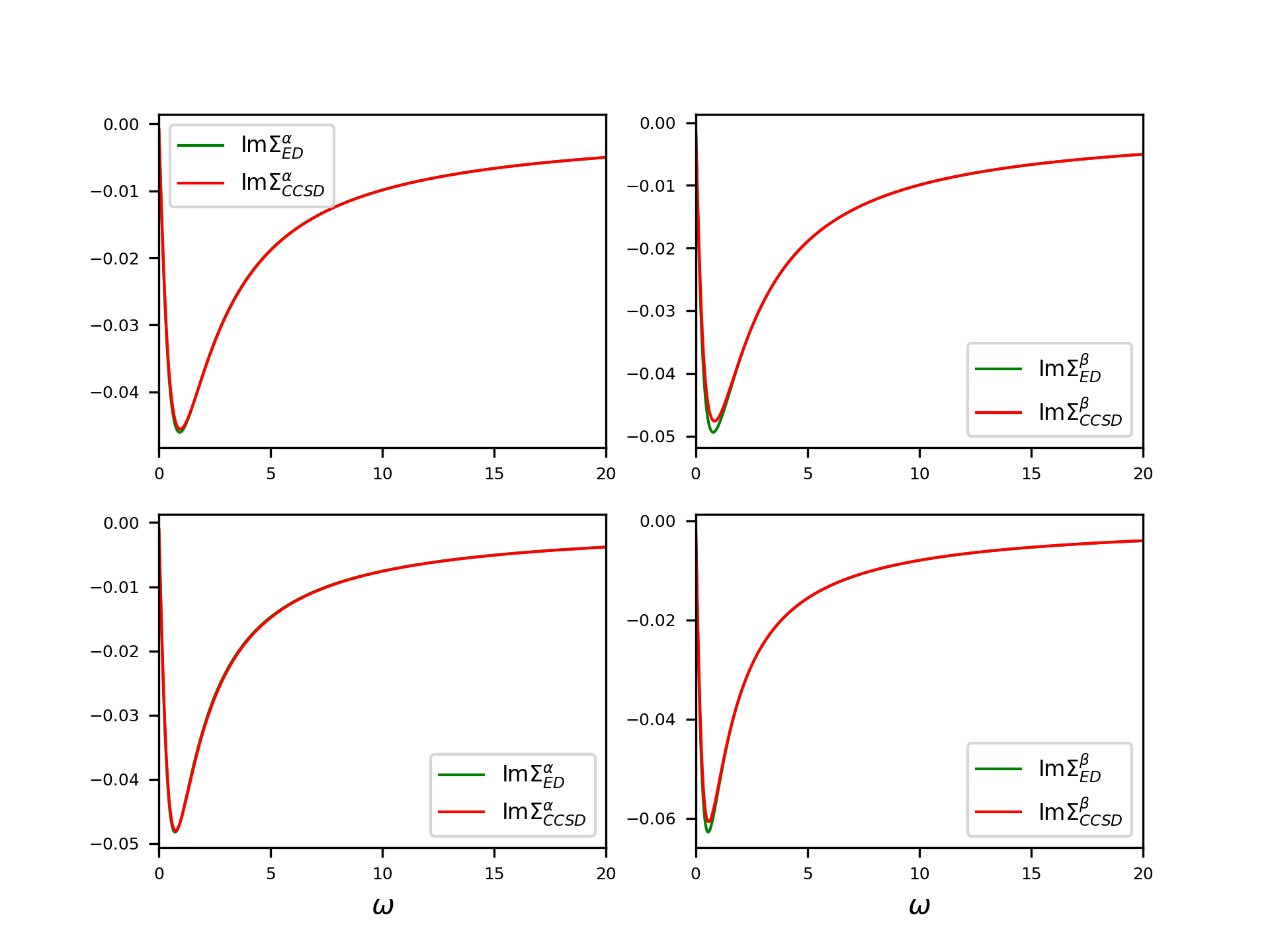} 
\end{center}
\caption{Self-energies are plotted for $\sigma_{Ti:3d_{z^2}+O:2p_z}$ (top panel) and $\sigma^*_{Ti:3d_{z^2}-O:2p_z}$ (bottom panel) orbitals of TiO molecule.}
\label{fig:TiOsigma2}
\end{figure}

We have applied SEET(CCSD/ED) method to a TiO molecule in the cc-pVDZ basis set. When we compared the total energy in Tab.~\ref{tab:CCSD/ED} against the previously obtained results, we observed that SEET(CCSD/ED) does not provide any significant improvement in comparison to the full system CCSD method. This result is quite surprising since the SEET(GF2/ED) method shows an improvement upon the GF2 method. Thus, it is natural to assume that the SEET(CCSD/ED) energy may show some improvement over the CCSD one and get closer to the SHCI energy. 
To investigate the source of this surprising observation, we have analyzed the self-energies for the impurity problems from Eq.~\ref{Eq:sigma_update_cc-nonloc}.
We observed that CCSD and ED methods yield almost identical self-energies  for the impurity problem containing many fewer  orbitals than the original system. We have plotted these self-energies in Fig.~\ref{fig:TiOsigma2} to demonstrate this behaviour. As a consequence, the total impurity self-energy becomes nearly zero. Therefore, from Eq.~\ref{Eq:sigma_update_cc-nonloc}, we get $\Sigma_{total} \approx \Sigma^{CCSD}_{weak}$ even after assembling corrections due to ED. 
While the similarity between the CCSD self-energy and ED self-energy for the impurity problem precludes 
getting a desired correction from the local method (here ED), this particular aspect is encouraging when CCSD is used as a solver for the impurity. 

\begin{table*}
\centering
\caption{{Total energy listed in a.u. and computed at various theoretical levels for TiO in the cc-pVDZ basis set. CCSD natural orbitals were used
as the impurity basis for SEET(CCSD/ED) calculations and GF2 natural orbitals for SEET(GF2/ED) calculations. Choice of impurity is listed in Tab. \ref{tab:impurity_choice}.}}
\label{tab:CCSD/ED}
\begin{tabular}{|c|c|c|c|c|c| }
\hline 
 method    & GF2 & CCSD (GM) & SEET(GF2/ED) \cite{MB_comparison:2020} & SEET(CCSD/ED) & SHCI \\
\hline
  E = & -73.8188  & -73.8622    & -73.9024  & -73.8635 & -73.9038 \\
 \hline 
\end{tabular}
\end{table*} 

\section{CCSD as an impurity solver} \label{sec:ccsd_local}
We have already explored CCSD as an impurity solver in the previous works both for molecules \cite{Shee2019} and real materials \cite{Yeh_SheePRB21}. Here we will summarize only the essential details. CCSD Green's function solver consists of two different parts: a) particle sector search ; b) CCSD Green's function construction. A particle sector search is essential because we define our local impurity problem using the Anderson impurity model (AIM), where the number of particles is not known \textit{a priori}. 
The particle search can be avoided when the number of particles present in the impurity is established based on the chemical potential given by  the non-local method which is set to be Fermi level of the impurity problem.
However, such a definition of Fermi level does not account for the change in the description of the correlation between the weak and strong correlation method and often leads to finding an erroneous particle sector \cite{Yeh_SheePRB21}. For this reason, we evaluated the minimum energy particle sector from the all possible distributions of particles within the number of orbitals chosen for the impurity problem. In the next step, we map the CC wave function at the singles doubles level (CCSD) to the corresponding Green's function using the Lehmann representation \cite{NooijenIJQC1992,KowalskiJCP2014,Shee2019}
\begin{align}
G^{CC}_{pq} (\omega) = {} & \langle \Phi | (1+ \Lambda) \overline{a_p^\dagger} \frac{1}{\omega + \mu + \overline{H} - i\eta } \overline{a}_q| \Phi \rangle \nonumber \\ 
     {} & +  \langle \Phi | (1+ \Lambda) \overline{a}_p \frac{1}{\omega + \mu - \overline{H} + i\eta}  \overline{a_q^\dagger} | \Phi \rangle \label{Eq:ccgf2},
\end{align}
where, $\overline{a}_p = e^{-T} a_p e^T$, $\overline{a_p^\dagger} = e^{-T} a_p^\dagger e^T$ with $a^{\dag}_{p}$ ($a_{p}$) being creation (annihilation) operators to the single-particle state $p$, and $\overline{H} = e^{-T} H e^T - E_{gr}$. $E_{gr}$ is the UCCSD ground state energy.

\begin{figure}
    \centering
    \includegraphics[width=0.55\textwidth]{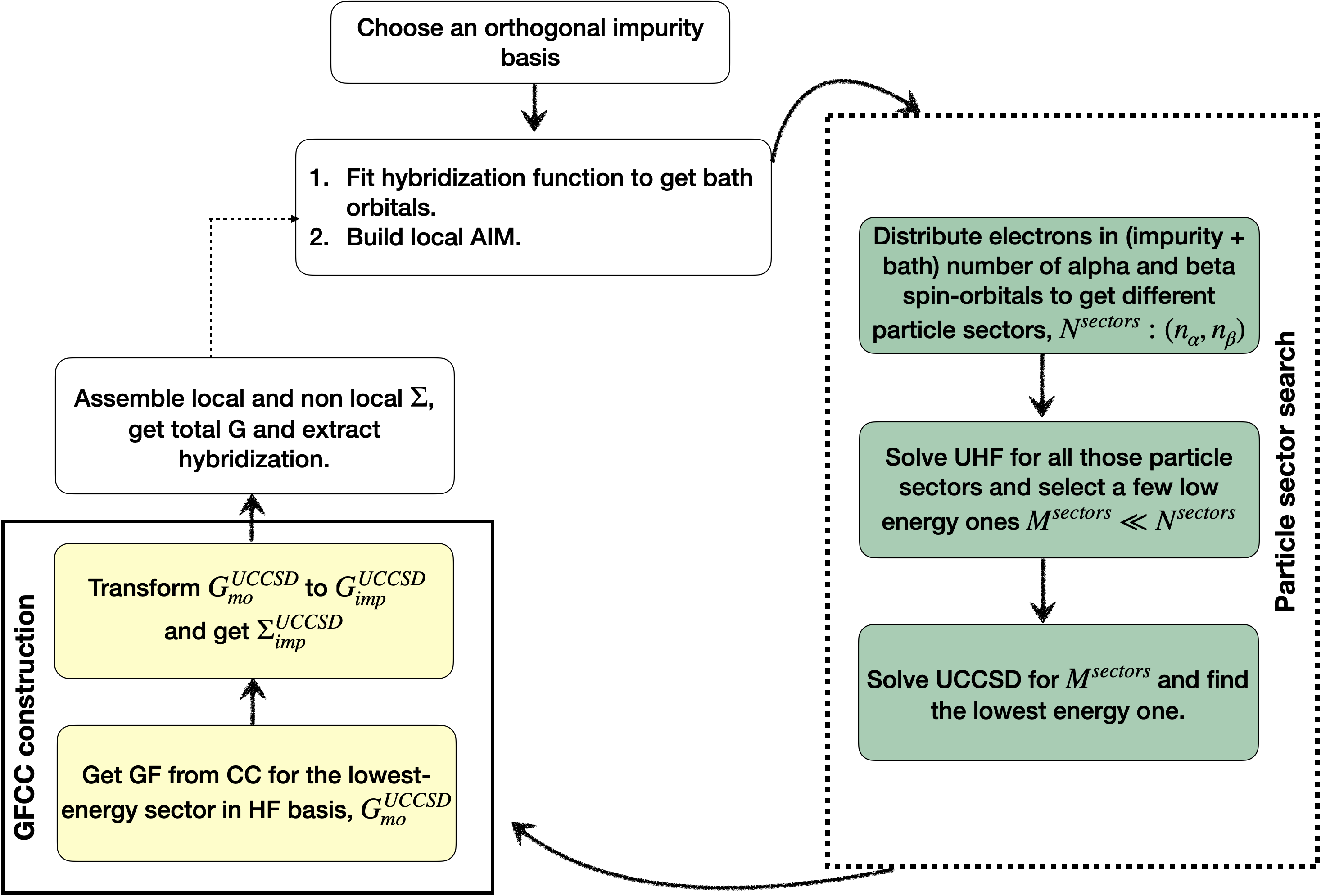}
    \caption{A schematic view of the inner loop of SEET using the GFCCSD solver. Particle sector search and GFCCSD construction sections constitute the solver section, and the rest of the steps are general to the SEET scheme.}
    \label{fig:ccsd_solver}
\end{figure}

In the computation of GFCCSD, we first tridiagonalize the Hamiltonian using the Lanczos method in the space of (N+1) and (N-1) excitations, and then evaluate the GF from a continued fraction formula. This particular way of evaluating GF has two advantages: i) the dominant cost of this approach is completely independent of the number of frequency points used; ii) computation does not suffer from the problem of instabilities for some of the frequency points as is the case for linear equation solver approach\cite{KowalskiJCP2014,NooijenIJQC1992}.    

\section{Computational Organization} \label{sec:organization}
When GFCCSD is used as a solver, we employ finite-temperature scGW evaluated on imaginary time and frequency grid as a non-local method. Since we want to evaluate the ground state total energy for all the molecules treated here, our calculations are carried out at very low-temperatures. A major challenging aspect of finite-temperature calculations is to choose suitable time and frequency grid that ensures high accuracy calculations.
In this regard, we have chosen the intermediate representation (IR) grid \cite{Yoshimi:IR:2017, IR_Gull_HiroshiPRB20}, where we only need a few hundred of $\tau$ and $\omega$ grid points to capture physics of a large energy scale and very low-temperature. In order to ensure low-temperature, we have used $\beta = 1000$ 1/a.u. for all the calculations.

We have used the cc-pVDZ basis set for all the calculations, and core orbitals are eliminated by using the effective core potential (ECP) of Trail \textit{et. al.} \cite{Trailecp15}. 
We also used the geometries of the molecules listed in Ref.~\citenum{MB_comparison:2020}.
All integral evaluations are carried out using the \texttt{PySCF} \cite{pyscf18} quantum chemistry package. Since our scGW code is written in terms of the 3-center 2-electron coulomb integrals \cite{Ren_2012, Iskakov20}, we use density fitted integrals \cite{dens_fit_pyscf17} from \texttt{PySCF}, where we choose even tempered Gaussians as the auxiliary basis. 

The integration of the GFCCSD solver with the SEET method is schematically depicted in Fig. \ref{fig:ccsd_solver}. Before entering the actual GFCCSD step, we ensure that we have found the right particle sector and converged unrestricted HF properly for that particle sector. For details of this procedure see our previous work \cite{Yeh_SheePRB21}. Also as depicted in Fig. \ref{fig:ccsd_solver}, we converge only inner loop of SEET.


As we use the Lanczos solver for GFCCSD, we have to use an arbitrary cut-off in the number of Lanczos chain vectors. We have established in our earlier work~\cite{Shee2019} that 100 vectors are sufficient since the results remain essentially unchanged even if we increase that number of vectors.


Another important aspect of our implementation is that we manage the evaluation of $O(N^2)$ elements of Green's function by efficiently parallelizing it via multiprocessing module of python, where $N$ is the total number of the orbitals in the impurity. The evaluation of GFCCSD for each of the elements of Green's function scales as $O(N^5)$.

\section{Choice of impurity basis}\label{Sec:imp_choice}
The choice of impurity basis for an embedding calculation is very important since final results depends crucially on that. In order to find a suitable basis for both the atomic and molecular examples studied here, we have started with symmetrized atomic orbital (SAO) basis. The choice of impurity orbitals was made based on the fact that d-orbitals of transition metals are strongly correlated, and from our chemical knowledge that O p-orbitals will form $\sigma$ and $\pi$ bonds with these d-orbitals. This means that our chosen impurity is composed of the metal 3d and 4d and O 2p and 3p orbitals. Furthermore, we have split these impurity orbitals according to their contribution to the $\sigma$ and $\pi$ bonding. 

The other choice of impurity orbitals is natural orbitals (NO) from earlier GW calculations. We analyze NOs obtained from GW quite thoroughly before selecting the suitable ones. First, we make a preliminary choice based on their occupation number to see which orbitals are strongly correlated. Afterwards, we find out the AOs they are composed of, which informs us about the precise bonding feature. For most of the molecules treated here, we use the same chemical bonding feature, as mentioned before, of metal d- and O p-, that is then subdivided into $\pi$ and $\sigma$ type bonding.

\begin{figure}
    \centering
    \includegraphics[width= 1 \linewidth, height=7cm]{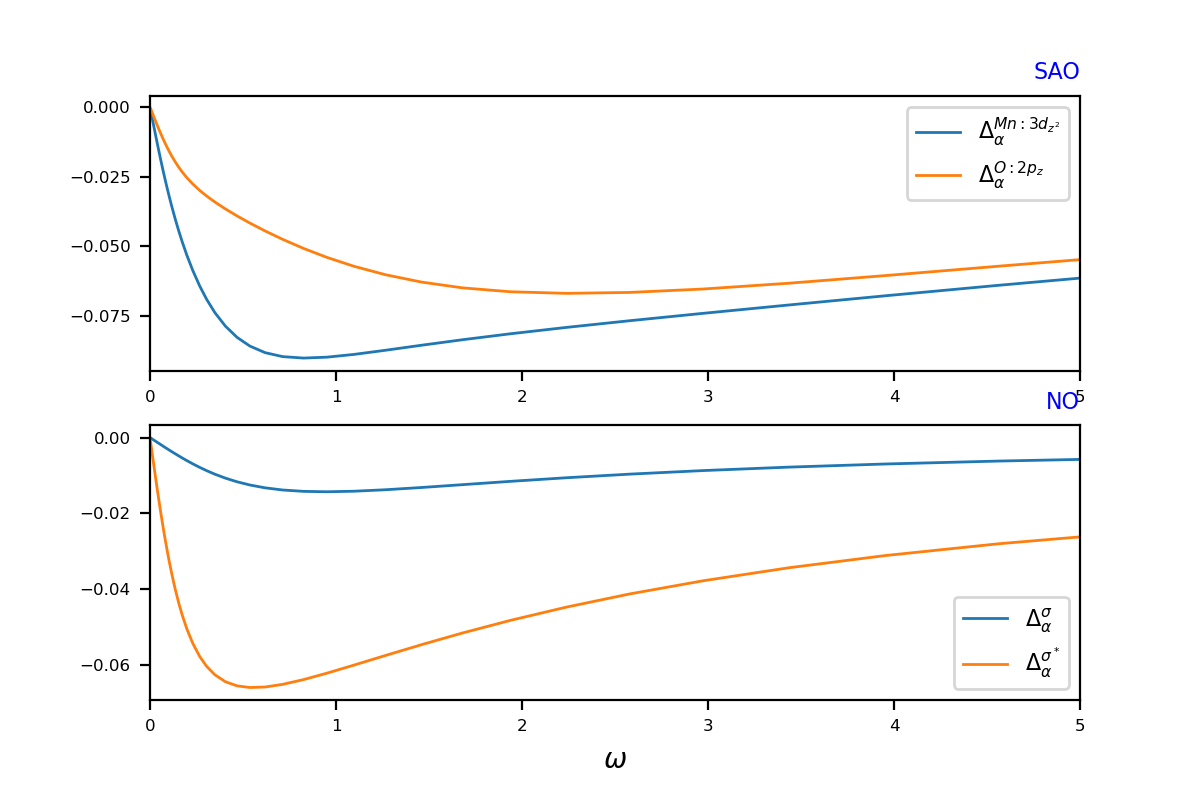}
    \caption{Comparison of hybridization elements evaluated in the SAO and NO impurity basis for ScO molecule.}
    \label{fig:sco_delta}
\end{figure}

In the following, we will present a comparison between these two choices. The example of ScO molecules is chosen for this comparison.  Where, with SAO, we have three groups of orbitals A: [Sc: 4s, 3d$_{z^2}$, 4d$_{z^2}$, O: 2p$_z$, 3p$_z$], B: [Sc: 3d$_{xz}$, 3d$_{yz}$, 4d$_{xz}$, 4d$_{yz}$, O: 2p$_x$, 2p$_y$, 3p$_x$, 3p$_y$], C: [3d$_{xy}$, 3d$_{x^2-y^2}$, 4d$_{xy}$, 4d$_{x^2-y^2}$].  On the other hand for NOs, the orbitals are grouped as A: [1$\sigma$, 1$\sigma^*$, 2$\sigma$, $\delta_{x^2-y^2}$], B:[$d_{\pi_x}$, $d_{\pi_y}$, $p_{\pi_x}$, $p_{\pi_y}$].
Our observations below are based on SEET(GW/CCSD) scheme where CCSD was used as an impurity solver while GW was used as a weakly correlated method for treating all orbitals. We want to emphasize here that this particular analysis is not possible with a solver such as ED, because the number of orbitals present in the impurity is too large. Here, we list our typical observations concerning the performance of impurity solvers using NO and SAO.
\begin{enumerate}
    \item We typically require fewer number of NOs to capture similar physical effects in comparison to SAO, because NOs are more compact combining several higher angular momentum AOs. This difference is quite apparent from the choice of impurities for ScO.
    \item Some hybridization elements are also much smaller in NO basis (figure \ref{fig:sco_delta}), 
    since the non-local effects are ``folded into'' the impurity due to the basis transform.
    However, one disadvantage of small hybridization elements is that bath fitting is often quite not as accurate as in SAO basis, where the hybridization elements are larger.
    \item Furthermore, our observation is that with NO we need fewer number of bath orbitals in the fitting scheme. The reason can be traced back to the small hybridization elements (see Fig. \ref{fig:sco_delta}), for which we either need very few or no bath orbitals to fit. For example, if we consider impurity A of both SAO type and NO type, where they consist of similar orbitals, the average number of bath orbitals for SAO type is  $> 5$ per impurity orbital, whereas for NO, the average number is $< 3$ per impurity orbital. This means that in effect, we generate much smaller impurity problem in NO basis. This is very advantageous from the computational point of view since such an impurity problem expressed in NO orbitals can then be treated with expensive impurity solvers.
    \item NOs, however, pose a challenge in terms of choosing a suitable set of orbitals for the impurity problem, whereas with SAO this choice is more unbiased, as we discussed before.
\end{enumerate}

Finally, when we compare total energies after the full convergence of iterative cycles in SEET, we get the energies listed in Tab.~\ref{tab:imp_basis}. Clearly, in the SAO basis, we fail to recover a good total energy close to the benchmark values for this system. We surmise that SAOs do not include enough of inter orbital correlations, hence we need quite a large number of them to recover a good total energy. Precisely for this reason for all other systems, we will be considering NOs as our choice of impurity basis.

\begin{table}
\caption{Comparison of total energy in a.u. for different choices of impurity basis for ScO molecule.}
\begin{tabular}{|c|c|c|}
\hline 
 SEET(GW/CCSD)-SAO & SEET(GW/CCSD)-NO & SHCI \\
\hline 
\hline 
 -62.473224 & -62.422761 & -62.42004 \\
\hline 
\end{tabular}\label{tab:imp_basis}
\end{table}

\section{Results} \label{sec:results}
\begin{figure}
    \centering
    \includegraphics[width= 1\linewidth, height=8cm]{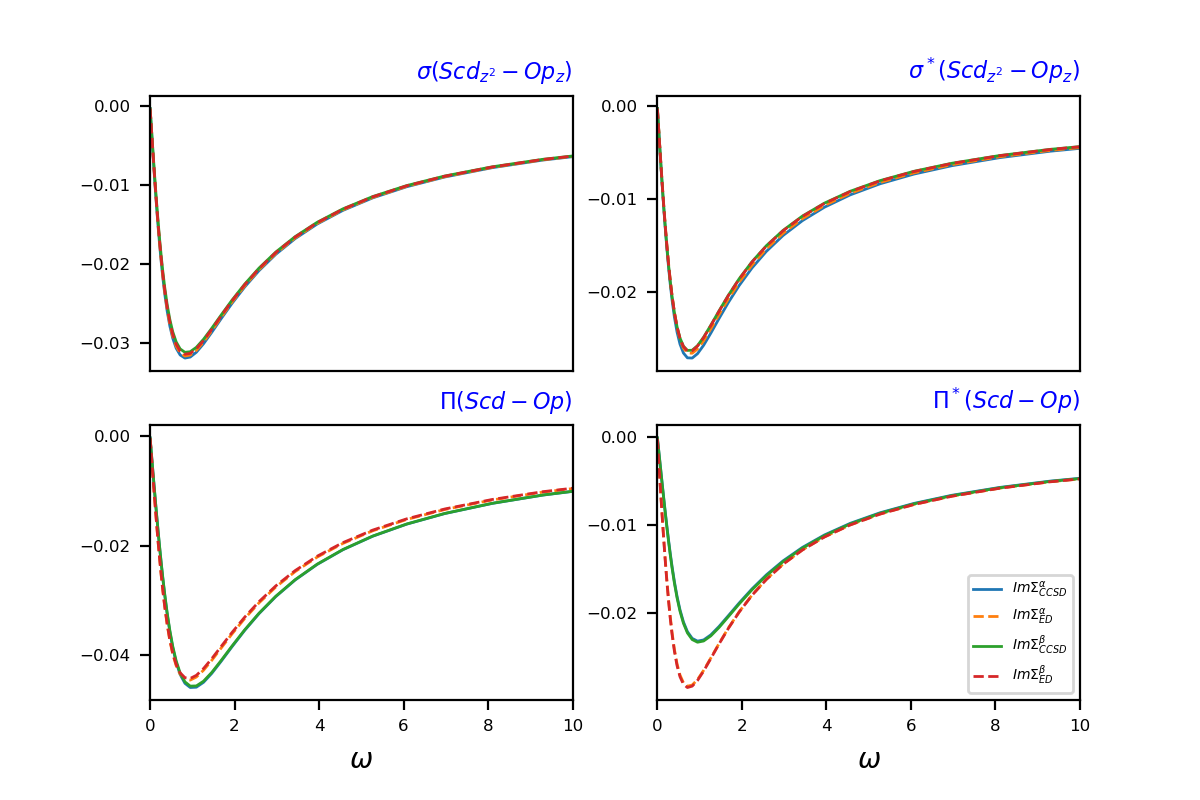}
    \caption{Comparison of ED and CCSD self-energies for ScO molecule. The self-energies of $\pi$, $\pi^*$ and $\sigma$, $\sigma^*$ orbitals are shown.}
    \label{fig:sco_sigma}
\end{figure}

The first molecule we considered was ScO.  For this case, we chose the impurity orbitals as described in Sec.~\ref{Sec:imp_choice}, that is, d$\pi$-p$\pi$ bonding and anti-bonding orbitals between Sc and O as one impurity and all the valence orbitals having major contribution from Sc: d$_{x^2-y^2}$ and d$_{z^2}$ orbitals as another impurity. This choice gives us two impurities, each of them containing 4 orbitals from ScO.  
Fitting of hybridization elements with discrete bath orbitals give 17 and 16 orbitals in total for the impurity problem, respectively. The impurity problems have been treated both with CCSD solver and ED solver. We observed that both these solvers found the same minimum energy particle sectors in the first iteration of SEET inner loop and the energy of the sectors are very much comparable - for impurity A they differ by 0.02 mH and for B by 5 mH. 
In Fig. \ref{fig:sco_sigma}, we have compared impurity self-energies from both solvers to observe that their behavior is consistent with that of total energy. We have plotted imaginary component of the Matsubara self-energy for the $\pi$ and $\sigma$ bonding and anti-bonding orbitals (the plotted self-energies are  obtained in the first iteration). Finally, the total energy of ScO, listed in Tab.~\ref{tab:all_oxides}, has been found to be very accurate with respect to the benchmark SHCI value. Interestingly, the total energy obtained from SEET(GW/CCSD) is even better than full system CCSD data - CCSD differs from SHCI by 32mH, and SEET(GW/CCSD) by 2.7mH.
This particular result is consistent with our observations made in the previous Sec.~\ref{sec:ccsd_nonloc}, where the SEET(GW/CCSD) energy was much better than the SEET(CCSD/ED) result that was comparable to the CCSD energy itself.
We think there could be two reasons behind this rather surprising outcome: i) accuracy of CCSD depends on the size of the impurity problem ;
ii) the choice of the impurity basis modifies the impurity problem such that even at the singles-doubles truncated level, we get almost converged  results (in terms of the rank of the cluster operators in the CC series). While we have not been able to carry out any rigorous numerical testing for assumption (ii), because it depends on the impurity choice, which is not unique and very hard to compare against any exact results, from theoretical CI studies \cite{MalrieuNO17}, we know that among many different set of orbitals, NOs provide one of the shortest expansions. Therefore, achieving a convergence at lower excitation rank is not surprising. 

In order to test the assumption (i), we have increased the size of the impurity problem by combining two impurities into one. With an increased impurity size, we do not have any higher-level methods for solving impurity problems that can yield very accurate self-energies. Therefore, we cannot easily compare CCSD self-energies against other methods. Instead, we have computed the total energy of the impurity problems with increasing rank of CC theory. The results are summarized in Tab.~\ref{tab:sco_analysis}. We observed that for impurity A, we get a converged total energy at the CCSD level, for impurity B improvement from CCSD to CCSDT is ~6 mH, which is substantial, and a reasonably converged total energy is obtained only at the CCSDTQ level. For A+B type impurity, we have 29 orbitals combining the impurity and bath orbitals. We can observe that convergence of the CCSD total energy for this case is worse than for any of the A and B impurities treated separately. Due to this reason the treatment of larger impurities (without adding additional excitation levels) may preclude obtaining even better total energies. We have anyway carried out the SEET self-consistency with A+B impurity as well, but total energy is ~18 mH below the SHCI benchmark number. Apart from the reason explained above, we also think it may happen because of the instability in finding the right particle sectors, as explained in our previous work~\cite{Yeh_SheePRB21}. We have not been able to carry out a thorough analysis in this regard as the size of the impurity is too large to handle using the ED solver.        

\begin{table}

\caption{Comparison of correlation energies in a.u. of various impurities with an increasing rank of CC theory for the ScO molecule. Impurity A, B and A+B contain 21, 20 and 29 orbitals, respectively. These numbers of orbitals are evaluated combining the orbitals from the system and bath.} \label{tab:sco_analysis}
\begin{tabular}{|c|c|c|c|c|}
\hline 
imp & CCSD & CCSDT & CCSDTQ & ED \\
\hline 
\hline 
A & -0.022722 & -0.022726 & -0.022726 & -0.022507 \\
\hline 
B & -0.058382 & -0.064160 & -0.063658 & -0.063602 \\
\hline 
A+B & -0.121790 & -0.135214 & -0.133416 &  too costly due to size\\
\hline 
\end{tabular}

\end{table}

\begin{figure}
    \centering
    \includegraphics[width= 1\linewidth, height=8cm]{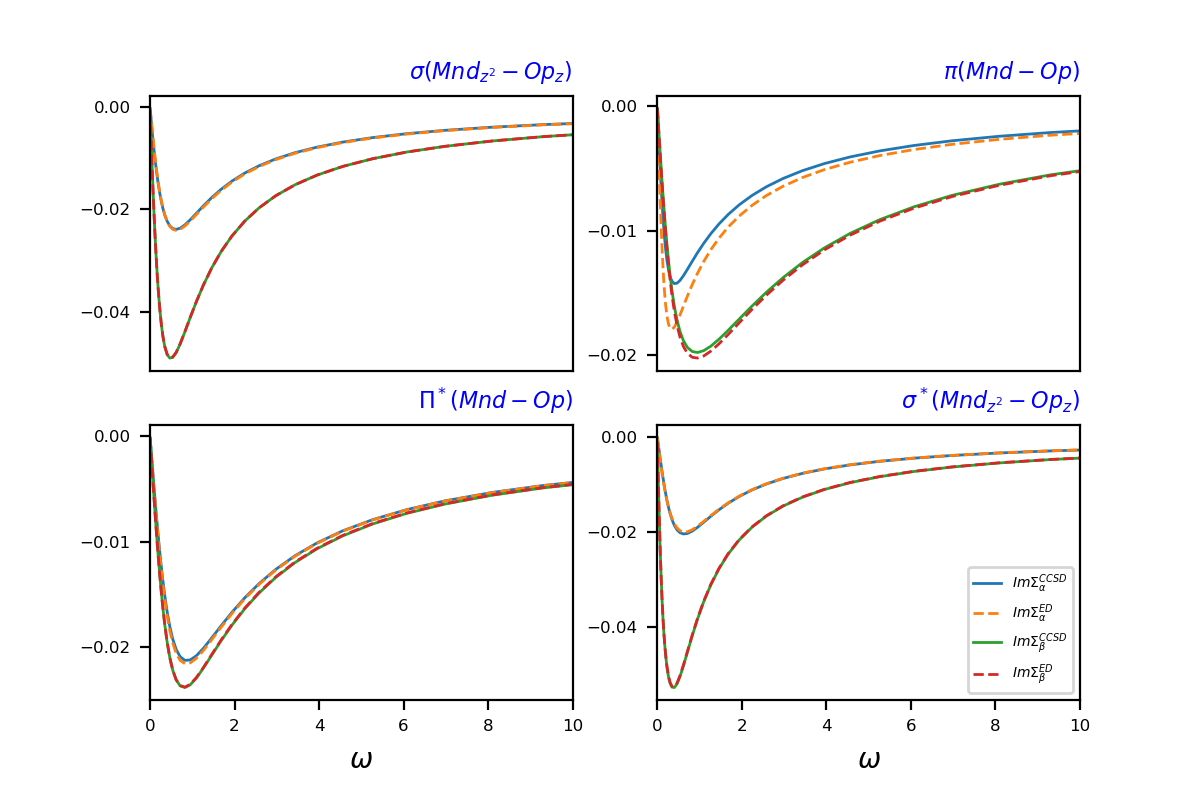}
    \caption{Comparison of the imaginary part of the Matsubara self-energies for MnO molecule. The self-energies of  $\sigma$, $\sigma^*$ and $\pi$, $\pi^*$ orbitals are shown.}
    \label{fig:mno_sigma}
\end{figure}

A subsequent system that we have considered is MnO. MnO has a $^6\Sigma^+$ ground state, and the distribution of unpaired electrons is: $\sigma^1 \pi^2 \delta^2$. In order to capture all the correlation effects, we have not only considered the valence $\sigma$, $\sigma^*$, $\pi$, $\pi^*$, $\delta$ orbitals, but also all the $\pi$ orbitals arising from the Mn:4d shell. The latter orbitals are deemed to be necessary to capture the ``double-shell" effect. Afterwards, we group the impurity orbitals in the following manner: ${[}\sigma, \sigma^*, \delta {]}$, ${[} 3d_{\pi_x}, 3p_{\pi_x}, 4d_{\pi_x}, 4p_{\pi_x}  {]}$, ${[} 3d_{\pi_y}, 3p_{\pi_y}, 4d_{\pi_y}, 4p_{\pi_y} {]}$.
In order to assess the quality of the CCSD solver, we have first compared the particle search procedure in the first iteration. The results are summarized in Tab.~\ref{tab:MnO_psearch}. We can see that CCSD and ED agree very well both in finding the ground state particle sector and their energies. We have then compared the self-energies from both methods in Fig.~\ref{fig:mno_sigma}. We have chosen self-energies corresponding to $\sigma$, $\sigma^*$, $\pi$ and $\pi^*$ orbitals for comparison, plotting only the imaginary part of the self-energy. We can see that self-energies for $\sigma$ and $\sigma^*$ and $\pi^*$ orbitals are very much comparable in ED and CCSD, but, $\alpha$ component of $\pi$ self-energy differ slightly at the low-frequency regime. Since in the case of MnO, the largest impurity problem contains 14 orbitals, it can be evaluated with the ED solver.  We have completed the full inner loop self-consistency in such a case. The obtained total SEET(GW/ED) energy is very much comparable with SEET(GW/CCSD) (see Tab.~\ref{tab:all_oxides}). Similar to the case of ScO, we observe in Tab.~\ref{tab:all_oxides} that SEET(GW/CCSD) provides much better agreement with SHCI than CCSD performed on the full system. We analyzed this result similarly as we did for ScO to see that smaller impurities have much better convergence in terms of total energy with the rank of CC hierarchy than the larger one. Larger impurity consists of all the orbitals from impurity A, B and C together. Our further attempt to achieve self-consistency with the larger impurity failed because of an unstable particle sector search. We have illustrated this claim in Tab.~\ref{tab:MnO_particle_search}. 

For the rest of the systems considered in this work, we will not go into a detailed analysis since most of our earlier observations and analysis hold for these systems as well. They only differ in the choice of the impurities. The chosen impurities are summarized in Tab.~\ref{tab:impurity_choice}. The key observations from all these studies are listed below:

\begin{enumerate}
    \item Assuming that SHCI gives a very close estimate of the total energy within a given basis set, scGW in all the cases overestimates the total energy. CCSD on the other hand underestimates the total energy by quite a big margin. CCSD(T) is in a reasonable proximity to the total energy within a given basis. This suggests that with superior dynamical correlation method, even within single reference framework, we can recover good total energies for linear transition metal oxides studied here.
    \item With the SEET(GW/CCSD) method, we were able to improve upon the GW method in all the cases. Total energies are within 2mH of SHCI results for all the transition metal oxides. These results are vastly better than full system CCSD of those molecules and even better than CCSD(T). 
    \item The choice of impurities are often problematic for these systems. In order to achieve very good total energies when using the Green's function CCSD solver, we have to keep individual impurities small. This happens due to the reasons we have explained for ScO and MnO. 
\end{enumerate}
    
\begin{table}

\caption{Energies obtained during a particle sector search procedure for MnO molecule, both with ED and CCSD. Particle sectors are denoted as $n_\alpha$ number of up spin and $n_\beta$ number of down spin electrons.} \label{tab:MnO_psearch}

\begin{tabular}{|c|c|c|c|c|}
\hline 
imp & \multicolumn{1}{c}{particle} & sector {($n_\alpha$, $n_\beta$)} & \multicolumn{1}{c}{Energy [a.u.]} & \\
\hline 
\hline 
 & CCSD & ED & CCSD & ED \\
\hline 
A & (5, 2) & (5, 2) & -0.009756 & -0.009768 \\
\hline 
B & (4, 2) & (4, 2) & -0.017697 & -0.017821 \\
\hline 
C & (4, 2) & (4, 2) & -0.017696 & -0.017867 \\
\hline 
\end{tabular}
\end{table}

\begin{table*}[]
\caption{
Comparison of correlation energies in a.u. for various impurities with an increasing rank of CC theory for the MnO molecule. 
For the combined impurity A+B+C, we were unable to carry out a CCSDTQ calculation. Separate impurities A, B, C, and a combined one A+B+C contain 10, 14, 14, and 47 orbitals in total, respectively.} \label{tab:MnO_analysis}
\begin{tabular}{|c|c|c|c|c|}
\hline 
imp & CCSD & CCSDT & CCSDTQ & ED \\
\hline 
\hline 
A & -0.009756 & -0.009761 & -0.009761 & -0.009768 \\
\hline 
B & -0.017697 & -0.017818 & -0.017818 & -0.017821 \\
\hline 
C & -0.017696 & -0.017818 & -0.017818 & -0.017867 \\
\hline 
A+B+C & -0.150818 & -0.167700 & too costly due to size & too costly due to size\\
\hline 
\end{tabular}\label{tab:MnO_particle_search}
\end{table*}

\begin{table*}[!ht]
\caption{Choice of impurities for transition metal oxides. In the last column, the total number of orbitals (impurity orbitals + bath orbitals) for each respective impurity from the middle column is listed.} \label{tab:impurity_choice}
\begin{tabular}{|c|c|c|}
\hline 
Molecule & Choice of impurities & Impurity (N$_{imp + bath}$) \\
\hline 
\hline 
ScO & A: {[} $\sigma_{Sc:3d_{z^2}+O:2p_z}$, $\sigma_{Sc:4s+3d_{z^2}+O:2p_z}$, $\sigma^*_{Sc:3d_z^2+O:2p_z}$, $\delta_{Sc:3d_{x^2-y^2}}${]}, & \\
& B: {[} $p_{\pi_x}$, $p_{\pi_y}$, $d_{\pi_x}$, $d_{\pi_y}$ {]} & A(17), B(16) \\
\hline 
TiO & A: {[} $\sigma_{Ti:3p_z+O:2s+3s}$, $\sigma_{Ti:3d_{z^2}+O:2p_z}$, $\delta_{Ti:3d_{xy}}$, $\sigma^*_{Ti:3d_{z^2}+O:2p_z}$ {]} & A(16) \\
\hline 
VO & A:{[}$\sigma_{V:3d_z^2+O:2p_z}$, $\sigma^*_{V:3d_z^2+O:2p_z}$, $\sigma_{V:3d_z^2+O:4p_z+4s}, \delta_{V:3d_{x^2-y^2}}${]}  & A(17)\\
\hline 
CrO & A:{[}$\sigma_{Cr:3d_z^2+O:2p_z}$, $\sigma_{Cr:4s+3d_z^2+O:2p_z}$, $\sigma^*_{Cr:4s+3d_z^2+O:2p_z}$, $\delta_{Cr:3d_{x^2-y^2}}${]} & A(17) \\
\hline
MnO& A: {[}$\sigma_{Mn: 3dz^2 + O: 2p_z}$, $\sigma^*_{Mn : 4s + 3dz^2 + O:2p_z}$, $\delta_{Mn: 3d_{xy}}$ {]}, & \\ & B: {[} $p\pi_{Mn: 3d_{xz} + O: 2px}$, $d\pi_{Mn: 3d_{xz} + O:2p_x}$, $\pi_{Mn:4p_x + O:4p_x}$, $d\pi_{Mn: 4d_{xz} + O: 4p_x}$ {]}, & \\
 & C: {[} $p\pi_{Mn: 3d_{yz} + p_y + O: 2p_y}$, $d\pi_{Mn: 3d_{yz} + O: 2p_y}$, $\pi_{Mn:4p_y + O:4p_y}$, $d\pi_{Mn: 4d_{yz} + O:4p_y}$  {]} & A(10); B(14); C(14) \\
\hline 
FeO & A:{[}$\sigma_{Fe:3d_z^2+O:2p_z}, \delta_{Fe:3d_x^2-y^2}${]} & A (10) \\
\hline 
CuO & A:{[}$\sigma_{Cu:3dz^2+ O:2p_z}$, $\sigma_{Cu:4s+O:2p_z}$, $\sigma^*_{Cu:4s+4dz^2+O:4s+p_z}${]}, B:{[}$d \pi_{Cu:3d_{yz}+O:2p_{y}+3p_{y}}$, & \\ 
& $\pi_{Cu:4p_x+O:2p_x+3p_x}$, $\pi_{Cu:4p_y+O:2p_y}$, $\pi^*_{Cu:4p_x+O:3p_x+4p_x}$, $\pi^*_{Cu:4p_y + 4d_{yz} + O:4p_y}$ {]} & A(11); B(22) \\
\hline 
\end{tabular}
\end{table*}

\begin{table*}[!ht]
    \centering
    \begin{tabular}{|c|c|c|c|c|c|c|}
      \hline 
      Molecule & GW & SEET(GW/CCSD) & SEET(GW/ED) & SHCI & CCSD & CCSD(T)\\
      \hline 
      \hline 
      ScO & -62.42865 & -62.42276  & N/P  & -62.42004 & -62.38804 & -62.41729 \\
      \hline 
      TiO & -73.91629 & -73.90397 & -73.90024 & -73.90375 & -73.86696 & -73.90072 \\
      \hline 
      VO & -87.09728 & -87.08300 & N/P & -87.08577 & -87.043325 & -87.08172 \\
      \hline 
      CrO & -102.57505 & -102.55919 & N/P & -102.55837 & -102.51743 & -102.55347 \\
      \hline 
      MnO & -119.87768 & -119.84873 & -119.84767 & -119.85051 & -119.81921 & -119.84550 \\
      \hline 
      FeO & -139.46151 & -139.43722 & -139.43971 & -139.43599 & -139.40078 & -139.43013 \\
      \hline 
      CuO & -213.16624 & -213.12640 & N/P & -213.12303 & -213.093983 & -213.11622 \\
      \hline 
    \end{tabular}
    \caption{Total energies in a.u. with various theoretical methods for all the transition metal oxides we have considered in this work.  For computational reasons, we have reported SEET(GW/ED) numbers only for cases where total size of the impurity is less than or equal to 16 orbitals. For TiO, the SEET(GW/ED) energy is reported only after the first iteration of the embedding loop. N/P stands for not performed due to the size of the Hilbert space present in the ED calculations (for more than 16 orbitals in the impurity problem).}
    \label{tab:all_oxides}
\end{table*}

\section{Discussion} \label{sec:concl}
In order for a SEET calculation to be predictive, we have to depend on several factors: a) accuracy of bath fitting; b) accuracy of the method used for describing weak correlation c)  choice of a suitable basis set for impurity; d) choice of the right orbitals for impurity; e) accuracy of the particle sector search and f) accuracy of the solver. Although it appears in the current work that we are interested only in the last factor (f), the final result is actually an interplay of all of them. For brevity reasons we will leave out (a) and (b) from the current discussion, except mentioning that they are quite satisfactory for the current work. We have discussed factor (c) in detail in Sec. \ref{Sec:imp_choice}, and opted for NOs from GW over SAOs. We have found that factors (d), (e) and (f) are completely intertwined, so in the following we will make general remarks based on all these three points.

The GFCCSD solver is quite comparable to ED solver, when the size of the impurity remains small. Therefore, if we can choose in a physically motivated way a suitable set of impurities containing a relatively small number of impurity orbitals, we can achieve quite good accuracy. Even though we have demonstrated it as an advantage, often times it can be quite crippling because it is not always obvious how to choose such an accurate set. It would have been more advantageous if we were able to increase the size of the impurity and achieve convergence with respect to the number of orbitals present in the impurity. GFCCSD allows us to increase the size, but its accuracy can deteriorate when larger impurities are considered. This, often leads to instability in particle sector search, and eventual divergence of SEET iterations. In addition, accuracy of the self-energy obtained from a larger impurities is diminished when compared to the smaller impurities. Therefore it is very hard to obtain systematic improvement.

The nature of the GFCCSD solver does not allow us to use the  generalized SEET scheme with overlapping impurities described in Ref.~\citenum{Tran_generalized_seet}. The reason is: generalized scheme relies on the fact that any double counting in self-energy between two different impurities are exactly cancelled. But, as already mentioned, for GFCCSD it is not guaranteed that the accuracy will be the same for all impurities. Therefore the exact cancellation may not happen.

The examples considered in this work point to the significance of the  triples correction. 
    In the future work, we will investigate if triples correction can provide the desired accuracy even with large impurities. It will allow us to check the convergence with respect to the CC hierarchy, thereby assess systematic improvement.  

\section{Acknowledgements}
A.S., Ch-N. Y., and D. Z. acknowledge support of the
Center for Scalable, Predictive methods for Excitation
and Correlated phenomena (SPEC), which is funded by
the U.S. Department of Energy (DOE), Office of Science,
Office of Basic Energy Sciences, the Division of Chemical
Sciences, Geosciences, and Biosciences.
\bibliographystyle{apsrev4-1}
\bibliography{cc, group, grid, misc, embed}


\end{document}